\documentclass[twocolumn,aps,prl,groupedaddress,showpacs]
              {revtex4}%
\usepackage{graphicx}
\usepackage{amsmath}
\usepackage{bm}
\usepackage{relsize}
\RequirePackage{xspace}
\begin{document}


\title{\mbox{}\\[10pt]
Next-to-Leading Order QCD Correction to $\bm{e^+ e^- \to J/\psi +
\eta_c}$ at $\sqrt {s}=10.6~$ GeV}

\author{Yu-Jie Zhang$~^{(a)}$, Ying-Jia Gao$~^{(a)}$, and Kuang-Ta Chao$~^{(b,a)}$}
\affiliation{ {\footnotesize (a)~Department of Physics, Peking
University,
 Beijing 100871, People's Republic of China}\\
{\footnotesize (b)~China Center of Advanced Science and Technology
(World Laboratory), Beijing 100080, People's Republic of China}}




\begin{abstract}
One of the most challenging open problems in heavy quarkonium
physics is the double charm production in $e^+e^-$ annihilation at
B factories. The measured cross section of $e^+ e^- \to J/\psi +
\eta_c$ is much larger than leading order (LO) theoretical
predictions. With the nonrelativistic QCD factorization formalism,
we calculate the next-to-leading order (NLO) QCD correction to
this process. Taking all one-loop self-energy, triangle, box, and
pentagon diagrams into account, and factoring the Coulomb-singular
term into the $c\bar c$ bound state wave function, we get an
ultraviolet and infrared finite correction to the cross section of
$e^+e^-\to J/\psi + \eta_c$ at $\sqrt{s} =10.6$~GeV. We find that
the NLO QCD correction can substantially enhance the cross section
with a K factor (the ratio of NLO to LO ) of about 1.8-2.1; hence
it greatly reduces the large discrepancy between theory and
experiment. With $m_c=1.4{\rm GeV}$ and $\mu=2m_c$, the NLO cross
section is estimated to be $18.9$~fb, which reaches to the lower
bound of experiment.

\end{abstract}


\maketitle

One of the most challenging open problems in heavy quarkonium
physics and nonrelativistic QCD (NRQCD) is the double charm
production in $e^+e^-$ annihilation at B factories. The inclusive
production cross section of $J/\psi$ via double $c\bar c$ in $e^+
e^- \!\to \!J/\psi c\bar c$ at $\sqrt{s}=10.6$GeV measured by
Belle Collaboration \cite{Abe:2002rb} is about a factor of 5
higher than theoretical predictions including both the
color-singlet\cite{cs} and color-octet\cite{liu04} $c\bar c$
contributions in the leading order (LO) NRQCD ~\cite{BBL}. Even
more seriously, the exclusive production cross section of double
charmonium in $e^+ e^- \!\to\! J/\psi \eta_c$  measured by Belle
\cite{Abe:2002rb, Pakhlov}
\begin{eqnarray}
\sigma[J/\psi + \eta_c] \times B^{\eta_c}[\ge 2]& =& \left( 25.6
\pm 2.8 \pm 3.4 \right) \; {\rm fb}, \label{Belle}
\end{eqnarray}
 and BaBar\cite{BaBar:2005}
\begin{eqnarray}
 \sigma[J/\psi + \eta_c] \times B^{\eta_c}[\ge 2]
&=& \left( 17.6 \pm 2.8 ^{+1.5}_{-2.1}\right) \; {\rm fb},
\label{BaBar}
\end{eqnarray}
could be larger than theoretical predictions by an order of
magnitude or at least a factor of 5. Here $B^{\eta_c}[\ge 2 ]$ is
the branching fraction for the $\eta_c$ to decay into at least 2
charged tracks, so Eqs.~(\ref{Belle}) and (\ref{BaBar}) give the
lower bound for this cross section. Theoretically, treating
charmonium as a nonrelativistic $c\bar c$ bound state, two
independent studies by Braaten and Lee ~\cite{Braaten:2002fi} and
by Liu, He, and Chao~\cite{Liu:2002wq} showed that at LO in the
QCD coupling constant $\alpha_s$ and the charm quark relative
velocity $v$ the cross-section of $e^+ e^- \to J/\psi \eta_c$ at
$\sqrt{s}=10.6$GeV is about $3.8\sim 5.5$fb (depending on the used
parameters, e.g., the long-distance matrix element, $m_c$ and
$\alpha_s$). In comparison with Eq.~(\ref{Belle}) or
Eq.~(\ref{BaBar}), such a large discrepancy between theory and
data may  present a challenge to our current understanding of
charmonium production based on NRQCD and perturbative QCD.

Some theoretical studies have been suggested in order to resolve
this large discrepancy problem. In particular, Bodwin, Braaten,
and Lee proposed~\cite{Bodwin:2002fk,Bodwin:2002kk} that processes
proceeding via two virtual photons may be important, and Belle
data for $J/\psi + \eta_c$ might essentially include the
$J/\psi+J/\psi$ events which were produced via two photons.
Brodsky, Goldhaber, and Lee suggested that since the dominant
mechanism for charmonium production in $e^+ e^-$ annihilation is
expected to be the color-singlet process $e^+e^- \to c \bar{c}
gg$, the final states observed by Belle might contain $J/\psi$ and
a $M \sim 3{\rm GeV}$ spin-$J$ glueball $\mathcal{G}_J$
($J=0,2$)~\cite{Brodsky:2003hv}. Motivated by these proposals,
Belle presented an updated analysis~\cite{Abe:2004ww}, and ruled
out the $J/\psi+J/\psi$ and spin-0 glueball scenarios. Ma and Si
studied this process by treating the charm quark as a light quark
and using light-cone distribution amplitudes to parameterize
nonperturbative effects related to the inner structure of
charmonium~\cite{MaandSi:2004}. Similar approaches were also
considered by Bondar and Chernyak~\cite{bondar}. But the enhanced
cross section is sensitive to the specific form of quark
distributions. Hagiwara, Kou and Qiao obtained a result consistent
with Ref.\cite{Braaten:2002fi} and Ref.\cite{Liu:2002wq}, and
conjectured that higher-order corrections in $\alpha_s$ may be
huge~\cite{Hagiwara:2003cw}. There are also other suggestions to
resolve the double charmonium problem, and a comprehensive review
on related topics and recent developments in quarkonium physics
can be found in Ref.~\cite{Brambilla:2004wf}.

 In order to further clarify
this problem, in this paper we present a result for  the next to
leading order (NLO) QCD correction to the process of $e^+ + e^- \to
J/\psi+\eta_c$. As is known, the NLO QCD corrections are important
for quarkonium production in inelastic $J/\psi$
photoproduction\cite{kra}, in $J/\psi$ plus jet and plus prompt
photon associated production in two photon collisions\cite{kni}, and
in gluon fragmentation functions for heavy quarkonium\cite{bra}.

\begin{figure}
\includegraphics[width=9.0cm]{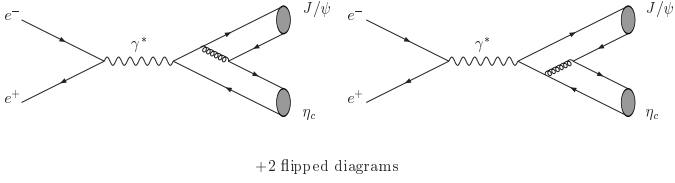}
\caption{\label{fig1} Born diagrams for $e^-(k_1) e^+ (k_2)\to
J/\psi(2p_1) \eta_c(2p_2)$. }
\end{figure}

At LO in $\alpha_s$, $J/\psi + \eta_c$ can be produced at order
$\alpha^2\alpha_s^2$, for which we refer to e.g.
Ref~\cite{Liu:2002wq}. There are four Feynman diagrams, two of which
are shown in Fig.~\ref{fig1}, and the other two can be obtained by
reversing the arrows on the quark lines. Momenta for the involved
particles are assigned as $e^-(k_1) e^+ (k_2)\to J/\psi (2p_1)+
\eta_c (2p_2)$. Using the NRQCD factorization formalism, we can
write down the scattering amplitude in the nonrelativistic limit to
describe the creation of two color-singlet $c \bar c$ pairs at short
distances, which subsequently hadronize into $J/\psi + \eta_c$ at
long distances in the $e^+e^-$ annihilation process. (Note that here
the color-octet $c\bar c$ contribution is of higher order in $v$ and
therefore negligible). Choosing the Feynman gauge, we get the
amplitude of Born diagrams
\begin{eqnarray}
i\mathcal{M}_{Born}&=& \frac{4096\pi e_c\alpha \alpha_s
m|R_S(0)|^2}{3s^3} \times
 \nonumber\\
&&  \epsilon_{\alpha \beta \nu \rho } p_{1}^{\alpha}p_{2}^{\beta}
\varepsilon^{* \nu} \bar{v}_e(k_2)\gamma^\rho u_e(k_1),
\label{eq:amp}
\end{eqnarray}
where $s=(k_1+k_2)^2$, $e_c=\frac{2}{3}$ is the electric charge of
the charm quark, $\rho$ is the Lorentz indices of the virtual
photon, $\varepsilon$ is the polarization vector of $J/\psi$.
$2p_1$ and $2p_2$ are the momenta of $J/\psi$ and $\eta_c$
respectively. $R_S(0)$ is the radial wave function at the origin
of the ground state charmonium $J/\psi$ and $\eta_c$.

\begin{figure*}
\includegraphics[width=14.5cm]{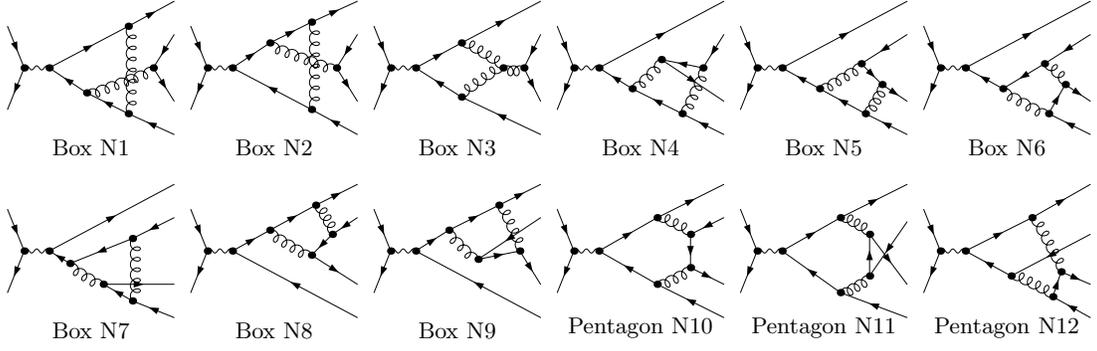}
\caption{\label{fig2} Twelve of the twenty-four box and pentagon
diagrams for $e^-(k_1) e^+ (k_2)\to J/\psi(2p_1) \eta_c(2p_2)$. }
\end{figure*}

At NLO in $\alpha_s$, the cross section is
\begin{eqnarray}
{\rm d}\sigma& \propto &|\mathcal{M}_{Born} + \mathcal{M}_{NLO}|^2 \nonumber \\
&=&|\mathcal{M}_{Born}|^2  \hspace{-0.1 cm}+ \hspace{-0.1 cm} 2
{\rm Re}(\mathcal{M}_{Born}\mathcal{M}_{NLO}^*) \hspace{-0.1 cm}+
\hspace{-0.1 cm} {\cal{O}}(\alpha^2 \alpha_s^4).
\end{eqnarray}
The self-energy and triangle diagrams all correspond to
propagators and vertexes of Born diagrams. There remain
twenty-four box and pentagon diagrams. Twelve diagrams of them are
shown in Fig.~\ref{fig2}. The upper $c \bar{c}$ hadronize to
$J/\psi$, and the lower to $\eta_c$. The other twelve diagrams are
obtained by reversing the arrows on the quark lines. Specially,
the associated diagram with Pentagon N12 exists only by reversing
the arrows on the lower quark lines which hadronize to $\eta_c$.

The self-energy and triangle diagrams are in general ultraviolet
(UV) divergent; while the triangle, box, and pentagon diagrams are
in general infrared (IR) divergent. Box N5 and N8 and Pentagon
N10, which have a virtual gluon line connected with the $c\bar{c}$
in a meson, also contain the Coulomb singularities due to the
exchange of longitudinal gluons  between $c$ and $\bar c$. In the
practical calculation, the IR and UV singularities are regularized
with $D=4-2\epsilon$ space-time dimension, and the Coulomb
singularities are regularized by a small relative velocity $v$
between $c$ and $\bar{c}$ \cite{kra}, $v=|\overrightarrow{p_{1c}}-
\overrightarrow{p_{1\bar c}}|/m$ , defined in the meson rest
frame. For the Coulomb-singular part of the virtual cross section,
we find
\begin{eqnarray}
\sigma &=& |R_S(0)|^4\hat\sigma^{(0)}\left(1
\hspace{-0.05cm}+\hspace{-0.05cm} \frac{2\pi
 \alpha_s C_{F}}{v}\hspace{-0.05cm} +\hspace{-0.05cm} \frac{\alpha_s\hat{C}}{\pi}
\hspace{-0.05cm}+\hspace{-0.05cm}{\cal{O}}(\alpha_s^2)\right)
\label{coulomb} \nonumber \\
&\Rightarrow & \hspace{-0.1cm} |R_S(0)|^4\;\hat
\sigma^{(0)}\left[1+\frac{\alpha_s}{\pi}\hat{C}+
{\cal{O}}(\alpha_s^2)\right]. \quad
\end{eqnarray}
In the second step, the Coulomb-singularity term has to be factored
out and mapped into the wave functions of $J/\psi$ and $\eta_c$. For
the LO expressions of operators $\left\langle{\cal
O}^{J/\psi}\left[{}^3\!S_1^{(1)}\right]\right\rangle$ and
$\left\langle{\cal
O}^{\eta_c}\left[{}^1\!S_0^{(1)}\right]\right\rangle$ are associated
with  $R_S(0)$, and the NLO are proportional to $ {\pi
 \alpha_s C_{F}/v}$ \cite{BBL}. And the two operators give a factor of $2$
 at ${\cal{O}}(\alpha_s)$ , resulting in just the
Coulomb-singular term in Eq.~(\ref{coulomb}).

The self-energy and triangle diagrams contain UV singularities,
which are removed by the renormalization of the QCD coupling
constant $g_s$, the charm-quark mass $m$ and field $\psi$, and the
gluon field $A_\mu$. Similar to the renormalization scheme in
Ref.\cite{kni}(see also \cite{kra}), we define
\begin{equation}
g_s^0\hspace{-0.1cm}=Z_gg_s  ,  \hspace{-0.6cm}\qquad
m^0\hspace{-0.1cm}=Z_mm  , \hspace{-0.6cm}\qquad
\psi^0\hspace{-0.1cm}=\sqrt{Z_2}\psi  , \hspace{-0.6cm}\qquad
A_\mu^0\hspace{-0.1cm}=\sqrt{Z_3}A_\mu,
\end{equation}
where the superscript 0 labels bare quantities and $Z_i=1+\delta
Z_i$, with $i=g,m,2,3$, are renormalization constants. The
quantities $\delta Z_i$ are of ${\cal O}(\alpha_s)$ and they
contain UV singularities and finite pieces which depend on the
choice of renormalization scheme. We define $Z_2$ and $Z_m$ in the
on-mass-shell (OS) scheme, and $Z_3$ and $Z_g$ in the modified
minimal-subtraction ($\overline{\rm MS}$) scheme
\begin{eqnarray}
\delta Z_2^{\rm OS}&=&-C_F\frac{\alpha_s}{4\pi}
\left[\frac{1}{\epsilon_{\rm UV}}+\frac{2}{\epsilon_{\rm IR}}
-3\gamma_E+3\ln\frac{4\pi\mu^2}{m^2}+4\right],
\nonumber\\
 \delta Z_m^{\rm OS}&=&-3C_F\frac{\alpha_s}{4\pi}
\left[\frac{1}{\epsilon_{\rm
UV}}-\gamma_E+\ln\frac{4\pi\mu^2}{m^2} +\frac{4}{3}\right], \nonumber\\
 \delta Z_3^{\overline{\rm
MS}}&=&\frac{\alpha_s}{4\pi} (\beta_0-2C_A)
\left[\frac{1}{\epsilon_{\rm UV}} -\gamma_E + \ln(4\pi)\right], \nonumber\\
  \delta Z_g^{\overline{\rm MS}}&=&-\frac{\beta_0}{2}\,
  \frac{\alpha_s}{4\pi}
  \left[\frac{1}{\epsilon_{\rm UV}} -\gamma_E + \ln(4\pi)
  \right],
\end{eqnarray}
where  $\mu$ is the renormalization scale, $\gamma_E$ is the Euler's
constant and $\beta_0=(11/3)C_A-(4/3)T_Fn_f$ is the one-loop
coefficient of the QCD beta function, and $n_f$ is the number of
active quark flavors. There are three  massless light quarks $u,d,s$
and one heavy quark $c$, so $n_f=4$.  Color factors are given by
$T_F=1/2,C_F=4/3,C_A=3$ in $SU(3)_c$. Differing from Ref.\cite{kni},
we take the $\overline{\rm MS}$ scheme for $Z_3$ with no external
gluon legs and set  $n_f=4$. In this scheme, we do not need to
calculate the self-energy on external quark legs. It turned out that
the difference for the calculated cross section in different schemes
is of order of next to next to leading order and can therefore be
neglected in the NLO result. In the NLO corrections we should use
the two-loop formula for $\alpha_s(\mu)$,
\begin{equation}
\frac{\alpha_s(\mu)}{4\pi}=\frac{1}{\beta_0L} -\frac{\beta_1\ln
L}{\beta_0^3L^2}, \label{eq:as}
\end{equation}
where $L=\ln\left(\mu^2/\Lambda_{\rm QCD}^2\right)$, and
$\beta_1=(34/3){C_A}^2-4 C_F T_F n_f-(20/3)C_A T_F n_f$ is the
two-loop coefficient of the QCD beta function.

Pentagon diagrams N11 and N12 can be reduced to integrals with a
lower number of external legs directly, since there are only two
independent momenta. Then they can be calculated the same way as
box diagrams. To treat Pentagon N10 in Fig.~\ref{fig2}, we need to
calculate the five-point function
$E_0[p_1,2p_1,-p_2,-2p_2,m,0,m,0,m]$, and the finite term
$E_0^{fin}$, where
\begin{widetext}
\begin{eqnarray}
\label{Ezero} E_0 &=&
E_0^{fin}+\frac{2}{s}D_0[-p_1,-p_1-p_2,p_1,0,m,0,m]+\frac{2}{s}D_0[p_2,p_1+p_2,-p_2,0,m,0,m],\\
E_0^{fin}&=&\frac{-4}{s}D_0[p_1\hspace{-0.1cm}+\hspace{-0.1cm}p_2,p_1\hspace{-0.1cm}+\hspace{-0.1cm}2p_2,-p_1,0,0,m,m]
+\hspace{-0.15cm}\int\hspace{-0.15cm} \frac{\mathrm{d}
^Dq}{(2\pi)^D} \frac{2/s(s/2- 4q\cdot p_1+ 4q \cdot p_2
-8m^2)}{(q^2\hspace{-0.1cm}-\hspace{-0.1cm}m^2)(q\hspace{-0.1cm}+\hspace{-0.1cm}p_1)^2
((q\hspace{-0.1cm}+\hspace{-0.1cm}2p_1)^2\hspace{-0.1cm}-\hspace{-0.1cm}m^2)
(q\hspace{-0.1cm}-\hspace{-0.1cm}p_2)^2((q\hspace{-0.1cm}-\hspace{-0.1cm}2p_2)^2
\hspace{-0.1cm}-\hspace{-0.1cm}m^2)}\nonumber \\
&=&\frac{ 2\sqrt{4m^2-s}\,\,
\mathrm{tan}^{-1}\frac{\sqrt{s}}{\sqrt{4m^2-s}}-
\sqrt{s}\,\,\mathrm{ln} (\frac{-s}{m^2})}{-i\pi^2 m^2s^{5/2}}
+\frac {2 (4m^2 \hspace{-0.1cm}-\hspace{-0.1cm} s)^{3/2}
\mathrm{tan}^ {\hspace{-0.0cm}-1}
 \hspace{-0.3cm}\frac {\sqrt s}
{\sqrt {4m^2-s}}\hspace{-0.1cm}+\hspace{-0.1cm} \sqrt s  \left(i
\pi
(3m^2\hspace{-0.1cm}-\hspace{-0.1cm}s)\hspace{-0.1cm}+\hspace{-0.1cm}(s\hspace{-0.1cm}-\hspace{-0.1cm}4m^2)\mathrm{ln}
({\frac {-s}{m^2}})\right)}
{8im^4\pi^2(4m^2-s)s^{5/2}(16m^2-s)^{-1}},\nonumber\\
\end{eqnarray}
\end{widetext}
where the IR- and Coulomb-finite term $E_0^{fin}$ is calculated
with dimension $D=4$ and velocity $v=0$, and
$\mathrm{ln}(-s/m^2)=\mathrm{ln}(-(s+i0)/m^2)=\mathrm{ln}(s/m^2)-i\pi$.
In Eq.~(\ref{Ezero}) the $D_0[-p_1,-p_1-p_2,p_1,0,m,0,m]$ term is,
\begin{eqnarray}
\hspace{-1cm}D_{0} &=&\frac{4}{s}C_0[-p_1,p_1,0,m,m]+ \frac{i}{(4
\pi)^2}\frac{2i\pi -2\mathrm{ln}4}{m^2s}.
\end{eqnarray}
This term will appear in $\mathrm{Box\, N5 ,~N8}$. The other
IR-divergence terms can be calculated like that. Then all the
IR-divergence terms become $C_0[p_1,-p_2,0,m,m]$ and
$C_0[p_{1c},-p_{1\bar{c}},0,m,m]$. We find that $\mathrm{Box\, N3
,N6,N7}$ and $\mathrm{Pentagon N12}$ are IR-finite respectively,
and sum of $\mathrm{Box\, N1+N2+N4+N9}$ is IR-finite, and
IR-divergence term of $\mathrm{Pentagon \,N11}$ is canceled by
vertex diagrams. IR-divergence and Coulomb-singular terms of
$\mathrm{Box \,N5+N8 }$ and $\mathrm{Pentagon \,N10} $ are all
related to the $C_0[p_{1c},-p_{1\bar{c}},0,m,m]$ term. With
$v=|\overrightarrow{p_{1c}}- \overrightarrow{p_{1\bar c}}|/m \to
0$,
\begin{eqnarray}
C_0= \frac{-i}{2m^2(4\pi)^2}\left(\frac{4 \pi
\mu^2}{m^2}\right)^{\epsilon}\Gamma(1+\epsilon)\left[\,
\frac{1}{\epsilon} + \frac{\pi^2}{v} -2 \right].
\end{eqnarray}
IR-divergence terms of $\mathrm{Box \,N5+N8 +Pentagon \,N10} $ are
canceled by counter terms, and the Coulomb singularity is mapped
into $R_S(0)$. UV term is canceled by counter terms. Then the
final NLO result for the cross section is UV-, IR-, and
Coulomb-finite. Details of the calculation can be found in a
forthcoming paper.

\begin{figure}
\includegraphics[width=8.0cm]{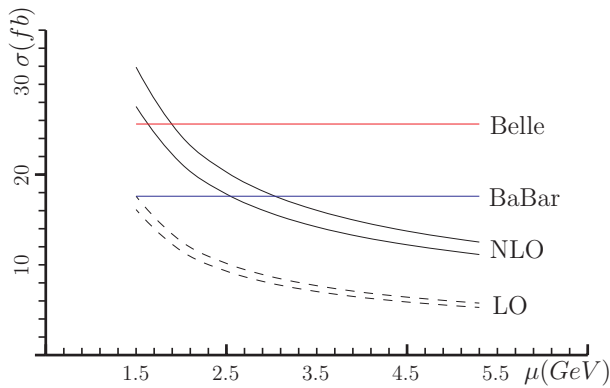}
\caption{\label{depmu}Cross sections as functions of the
renormalization scale $\mu$. Here $|R_S(0)|^2=0.978 {\rm GeV}^3$,
$\Lambda=0.338 {\rm GeV}$, $ \sqrt s=10.6{\rm GeV} $; NLO results
are represented by solid lines and LO one by dashed lines; the
upper line is for $m=1.4{\rm GeV}$ and the corresponding lower
line is for $ m=1.5{\rm GeV} $; the upper straight line denotes
the central value measured by Belle in Eq.(\ref{Belle}) and the
lower straight line by BaBar in Eq.(\ref{BaBar}).}
\end{figure}

We now turn into numerical calculations for the cross section of
$e^+ + e^-\rightarrow J/\psi+\eta_c$. To be consistent with the
NLO result the value of the wave function squared at the origin
should be extracted from the leptonic width at NLO of $\alpha_s$
(see e.g.~\cite{BBL}): $
|R_S(0)|^2=[(9m^2_{J/\psi})/(16\alpha^2(1-4 C_F \alpha_s/
\pi))]\Gamma(J/\psi\to e^+e^-) \label{eq:14} $. Using the
experimental value $5.40\pm 0.15\pm 0.07$~KeV~\cite{PDG}, we
obtain $|R_S(0)|^2=0.978 \mbox{GeV}^3$, which is a factor of 1.21
larger than $0.810 \mbox{GeV}^3$ that was used in
Refs.\cite{Braaten:2002fi,Liu:2002wq} from potential model
calculations. Taking $m_{J/\psi}=m_{\eta_c}=2m$ (in the
nonrelativistic limit),
 $m\!=\!1.5~GeV$, $\Lambda^{(4)}_{\overline{\rm MS}}=338{\rm MeV}$, with
Eq.(\ref{eq:as}) we find $\alpha_s(\mu)=0.259$ for $\mu=2m$ (these
are the same as in Ref.\cite{Liu:2002wq}, except here a larger
$|R_S(0)|^2$ is used), and get the cross section in NLO
\begin{equation}
\label{jsetac} \sigma(e^+ + e^-\rightarrow J/\psi+\eta_c)=15.7
\rm{fb},
\end{equation}
which is a factor of $1.96$ larger than the LO cross section
$8.0$~fb. If we set $\mu=m$ and $\mu=\sqrt{s}/2$, then
$\alpha_s=0.369$ and $ 0.211$, which result in the cross section
$27.5$ and $11.2$~fb respectively. If we set $m=1.4{\rm GeV}$ and
$\mu=2m$, the cross section is $18.9$~fb and $9.2$~fb at NLO and
LO respectively. (Our LO result is also consistent with
Ref.\cite{Braaten:2002fi} if we take their smaller value for
$|R_S(0)|^2$ and $\mu=\sqrt{s}/2$.)
In Fig.~\ref{depmu} we show the calculated $e^+ + e^-\rightarrow
J/\psi+\eta_c$ cross sections at LO and NLO as functions of the
renormalization scale $\mu$ with two mass values $m_c=1.4~$GeV and
$1.5~$GeV, as compared with the Belle and BaBar data. We see the
NLO QCD correction enhances the cross section by about a factor of
2, despite of existing theoretical uncertainties.

The relativistic corrections may further significantly enhance the
cross section\cite{HFC} (see also \cite{Braaten:2002fi}). The
reason for the enhancement is quite obvious that in
Fig.~\ref{fig1} the virtuality of the gluon takes its maximum
value of $Q^2=s/4$ in the nonrelativistic limit, and taking
account of the relative momentum between the charm quarks in the
charmonium will lower the value of the gluon virtuality.

In conclusion, we find that by taking all one-loop self energy,
triangle, box, and pentagon diagrams into account, and factoring
the Coulomb singular term associated with the exchange of
longitudinal gluons between $c$ and $\bar c$ into the $c\bar c$
bound state wave function, we get an ultraviolet (UV) and infrared
(IR) finite correction to the cross section of $e^+e^-\to J/\psi +
\eta_c$ at $\sqrt{s}=10.6$~GeV, and that the NLO QCD correction
can substantially enhance the cross section with a K factor (the
ratio of NLO to LO ) of about 1.8-2.1; and hence it crucially
reduces the large discrepancy between theory and experiment. With
$m=1.4{\rm GeV}$ and $\mu=2m$, the NLO cross section is estimated
to be $18.9$~fb, which reaches to the lower bound of experiment.

\begin{acknowledgments}
We would like to thank G.T. Bodwin, B.A. Kniehl  and J. Lee for
helpful comments and discussions. We also thank K.Y. Liu and C.
Meng for valuable discussions. This work was supported in part by
the National Natural Science Foundation of China (No 10421503),
the Key Grant Project of Chinese Ministry of Education (No
305001), and the Research Found for Doctorial Program of Higher
Education of China.
\end{acknowledgments}


\end{document}